\documentclass[aps,prl,twocolumn,superscriptaddress]{revtex4-2}
\usepackage{babel}
\usepackage{bbm}
\usepackage{mathrsfs}
\usepackage{amsmath}
\usepackage{amsfonts}
\usepackage[colorlinks=true,linkcolor=blue,urlcolor=blue,citecolor=blue,anchorcolor=blue]{hyperref}
\usepackage{graphicx}
\usepackage{subfigure}
\usepackage{epsfig}
\usepackage{dcolumn}
\usepackage{bm}
\usepackage{color}
\usepackage{natbib}
\usepackage{amssymb}
\usepackage{xcolor}
\usepackage{braket}
\usepackage{comment}

\usepackage{float}
\usepackage{lipsum}
\usepackage{pifont}
\usepackage[normalem]{ulem}

\definecolor{npurple}{rgb}{0.6, 0.3, 0.8}

\begin{document}

\title{Enhanced quantum metrology by criticality-assisted noncommutative preparation}

\author{Ningxin Kong}
\affiliation{\mbox{State Key Laboratory for Mesoscopic Physics, School of Physics, Frontiers Science Center for Nano-optoelectronics,} $\&$ Collaborative Innovation Center of Quantum Matter, Peking University, Beijing 100871, China}
\author{Matteo G. A. Paris}
\affiliation{Dipartimento di Fisica, Universit\`a di Milano, I-20132 Milano, Italy}

\author{Qiongyi He}
\email{qiongyihe@pku.edu.cn}
\affiliation{\mbox{State Key Laboratory for Mesoscopic Physics, School of Physics, Frontiers Science Center for Nano-optoelectronics,} $\&$ Collaborative Innovation Center of Quantum Matter, Peking University, Beijing 100871, China}
 \affiliation{Hefei National Laboratory, Hefei 230088, China}
\affiliation{\mbox{Collaborative Innovation Center of Extreme Optics, Shanxi University, Taiyuan, Shanxi 030006, China}}

\begin{abstract} 
Quantum criticality is a resource for quantum-enhanced metrology, but existing schemes face intrinsic limitations. These arise because using criticality directly in the encoding dynamics restricts  the accessible parameters to those explicitly supported by the critical Hamiltonian, and the requirement for critical conditions narrows the effective estimation range. To solve this, we introduce a general framework termed criticality-assisted noncommutative preparation (CANP). In this approach, critical evolution is employed as a state-preparation resource. We establish the underlying algebraic conditions and show that the intrinsic noncommutativity between the preparation and encoding operations leads to a genuine enhancement of the quantum Fisher information (QFI). Remarkably, this enhancement may be achieved at fixed total sensing time and energy cost. The effect is quantified by the Wigner–Yanase skew information, which measures noncommutativity and exhibits the same critical scaling as the QFI. We demonstrate effective use of CANP in the quantum Rabi and Lipkin-Meshkov-Glick models. Our results establish CANP as a robust technique to effectively implement criticality-enhanced quantum metrology.
\end{abstract}

    \maketitle

\textit{Introduction---}In quantum metrology, precise estimation of relevant parameters is achieved
by exploiting inherently quantum features as resources~\cite{Maccone2004, Maccone2006, Romalis2007, PARIS2009, Cappellaro2017, Treutlein2018}. Beyond superposition and entanglement~\cite{Maccone2011, Mitchell2011, Zoller2024, Zelevinsky2024}, quantum criticality has recently emerged as a  powerful resource for sensing, owing to the extreme sensitivity of systems near a critical point to small variations in physical parameters~\cite{Nikola2006, Lorenzo2008, Kehrein2013, Paris2016, Tommaso2018, Nigel2019,Bayat2025,Zeng2025}. Criticality has been so far harnessed to probe quantities that are directly linked to the proximity of systems to the critical point~\cite{Simone2020, Cai2021, Abolfazl2021, Karol2024, Candia2024}.
While such strategies can indeed yield significant enhancements, they are subject to inherent limitations~\cite{Jakub2018}. The set of estimable parameters is restricted by the structure of the critical Hamiltonian itself, typically limited to quantities such as coupling strengths or mode frequencies. Moreover, the requirement to operate at or near criticality narrows the accessible estimation range. These constraints fundamentally limit the versatility and broader applicability of criticality-enhance quantum metrology.

Beyond criticality, noncommutativity among quantum operations has been recognized as another important resource to enhance metrological precision~\cite{Giulio2020, Guo2023, Zeng2024, Wang2025}. Noncommutative dynamics 
may be exploited to improve the encoding of parameters, such as to reorganize the metrological protocol in a fundamentally more efficient manner, enabling sensitivity enhancements that are fundamentally inaccessible to commuting operations~\cite{Paris2025,He2026,Zeng2025b}. Despite these advances, the interplay between criticality and noncommutativity remains largely unexplored. 

Here, we introduce a technique, termed criticality-assisted noncommutative preparation protocol (CANP), 
in which the probe state is first prepared by a critical unitary evolution $\hat{U}_c$, and subsequently undergoes parameter encoding via unitary $\hat{U}_\theta$. In contrast to critically-based 
metrology protocols, our approach harnesses critical evolution as a state-preparation resource. We develop a general analytical framework and identify the algebraic conditions required to make 
criticality-assisted enhancement successful. Indeed, the noncommutativity between the prepared state and encoding Hamiltonians yields a substantial enhancement of the QFI, enabling improved sensitivity without increasing the total sensing time or energy cost. This effect is quantitatively captured by the Wigner–Yanase skew information, which faithfully tracks the noncommutative structure of the protocol. This quantity exhibits the same behavior as the QFI, both in the period of oscillations and in the location of the maxima. 
We illustrate the improvement in precision using the example of frequency estimation in the quantum Rabi model. We also show that realistic quadrature measurements can be employed to achieve performance close to the CANP-enhanced quantum Cram\`er–Rao bound. Overall, we show that relocating the critical Hamiltonian from encoding to  preparation naturally enlarges the class of accessible parameters beyond those contained in the original critical Hamiltonian and lifts the restriction of a narrow effective estimation range associated with critical conditions, thus broadening the scope of critical metrology.

\begin{figure}
    \centering
    \includegraphics[width=1\linewidth]{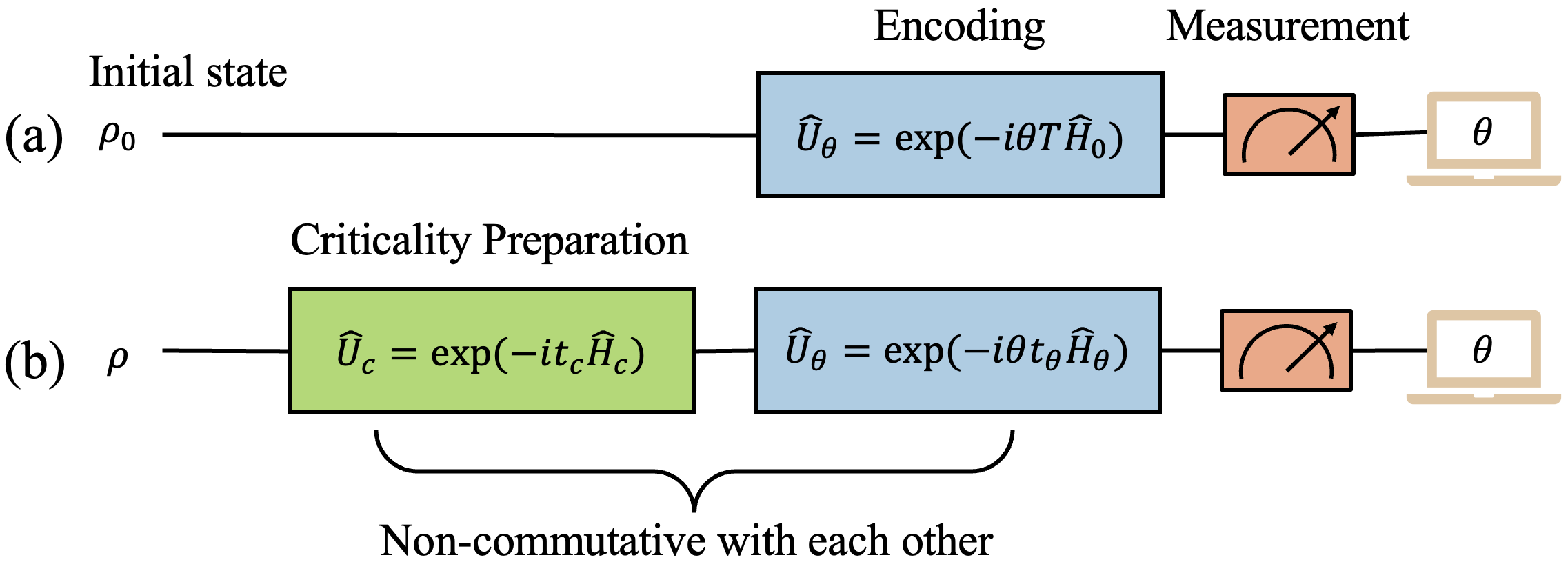}
    \caption{Schematic diagram of the quantum metrological protocols. (a) In 
    the direct-encoding scheme, a single probe state $\rho_0$ is directly subjected to the parameter-encoding operation $\hat{U}_\theta$ for a total encoding time $T$. (b) In our CANP protocol, the probe $\rho$ is first prepared by the critical operation $\hat{U}_c$ and subsequently acquires a dependence on the parameter $\theta$ through $\hat{U}_\theta$. The preparation and encoding durations are denoted by $t_c$ and $t_\theta$, respectively, with the total evolution time satisfying $T = t_c + t_\theta$.}
    \label{fig1}
\end{figure}

\textit{The CANP protocol---}In general, a quantum metrological protocol consists of four parts: the preparation of a probe state $\rho_0$, the encoding of an unknown parameter $\theta$ onto the probe, the measurement of an observable on the parameterized state $\rho_0(\theta)$, and the final data processing to extract the estimated value of $\theta$.
In a direct-encoding scheme [see Fig.~\ref{fig1}(a)], the parameter $\theta$ is imprinted through a unitary evolution $\hat{U}_\theta = \text{exp} (-i \theta t_\theta \hat{H}_\theta)$, where $\hat{H}_\theta$ is the Hermitian operator associated with $\theta$, and $t_\theta$ denotes the duration of the encoding stage. For a pure probe $\rho_0$, the QFI quantifying the distinguishability between neighboring parameterized states is given by $\mathcal{F}_0 (\theta) = 4\text{Var}[\hat{h}_0]_{\rho_0}$, where $\hat{h}_0 = i\hat{U}_\theta^\dagger \partial_\theta \hat{U}_\theta$ is the local generator of the encoding process, and $\text{Var}[\cdot]_{\rho_0}$ denotes the variance with respect to the probe state $\rho_0$~\cite{Caves1994, PARIS2009, Kok2010}. 

We now introduce the general framework of the CANP protocol, illustrated in Fig.~\ref{fig1}(b). For a given encoding Hamiltonian $\hat{H}_\theta$ associated with the unknown parameter $\theta$, the initial state $\rho$ is first evolved under a preparation Hamiltonian $\hat{H}_c$ that features quantum criticality. The preparation dynamics are required to satisfy the noncommutative algebraic relation
\begin{equation}
    [\hat{H}_c, \hat{\Gamma}] = \sqrt{\Delta} \hat{\Gamma},
    \label{eq1}
\end{equation}
where $\hat{\Gamma} = -i\sqrt{\Delta} \hat{C} + \hat{D}$ with $\hat{C} = i[\hat{H}_c, \hat{H}_\theta]$ and $\hat{D} = [\hat{H}_c, [\hat{H}_c, \hat{H}_\theta]]$. The critical parameter $\Delta$, determined solely by the intrinsic properties of the critical Hamiltonian $\hat{H}_c$, quantifies the proximity to the critical point. 
Importantly, the unknown parameter $\theta$ does not, in principle, enter either the critical Hamiltonian $\hat{H}_c$ or the criticality parameter $\Delta$. As a consequence, operating the system near the critical point does not restrict the accessible estimation range of $\theta$. This structural independence allows the CANP protocol to bypass the conventional trade-off in critical metrology, where the parameter of interest is directly tied to the critical behavior of the system.

This algebraic structure enables a closed-form expression for the local generator of parameter translations, $\hat{h} = i\hat{U}_c^\dagger \hat{U}_\theta^\dagger \partial_\theta \hat{U}_\theta \hat{U}_c$, which evaluates to:
\begin{equation}
    \hat{h} = t_\theta (\hat{H}_\theta + \frac{\sin(\sqrt{\Delta}t_c)}{\sqrt{\Delta}} \hat{C} + \frac{\cos(\sqrt{\Delta}t_c)-1}{\Delta} \hat{D}).
    \label{eq2}
\end{equation}
For preparation times $t_c \sim \mathcal{O}(1)$, the generator exhibits a significant amplification as $\Delta \to 0$, reflecting the emergence of critical dynamics during the preparation stage.
From Eq.~(\ref{eq2}), the corresponding QFI is given by
\begin{equation}
    \mathcal{F}(\theta) \simeq 4t_\theta^2 \frac{[\cos (\sqrt{\Delta} t_c)-1]^2}{\Delta^2} \text{Var}[\hat D]_{\rho}.
    \label{eq3}
\end{equation}
In the regime $\Delta \to 0$ with a finite preparation time $t_c \sim \mathcal{O}(1)$, it yields the scaling $\mathcal{F}(\theta) \sim t_\theta^2 t_c^4$, revealing a sensitivity enhancement at finite $t_c$ compared with the conventional $t_\theta^2$ behavior and enabling accelerate precision growth without prolong critical evolution. By contrast, when the preparation time becomes sufficiently long, $t_c \sim \pi/\sqrt{\Delta}$, the leading contribution leads to the divergent scaling $\mathcal{F}(\theta) \sim 16 \Delta^{-2} t_\theta^2$, reflecting a critical enhancement upon approaching the critical point. Importantly, these enhancements arise from the noncommutative structure of the protocol and persist for any initial state $\rho$ satisfying $\text{Var}[\hat{D}]_{\rho} \simeq \mathcal{O}(1)$, thereby eliminating the need for ground state preparation.

To assess the metrological gain arising from the CANP under fair resource constraints, we evaluate the QFI enhancement ratio $\mathcal{R}(\theta) = \mathcal{F}(\theta)/\mathcal{F}_0(\theta)$ while fixing both the available energy and the total sensing time. Without loss of generality, we consider the frequency estimation with $\hat{H}_\theta = \hat{a}^\dagger \hat{a}$, and take the probe to be a coherent state, $\rho = |\alpha\rangle \langle \alpha|$ [$\rho_0 = |\alpha_0\rangle \langle \alpha_0|$]. The quantities $\mathcal{F}(\theta)$ and $\mathcal{F}_0(\theta)$ denote the corresponding QFIs with or without a critical-preparation step, respectively. 
For the conventional directly-encoding scheme, the QFI simplifies to $\mathcal{F}_0(\theta) = 4 T^2 |\alpha_0|^2$, independent of $\theta$. To ensure a fair comparison between the two schemes, we require that the two final state, whether or not the critical-preparation is applied, carry the same average energy, $\langle \alpha_0 | \Omega \hat{a}^\dagger \hat{a}|\alpha_0\rangle  = \langle \alpha | (\hat{U}_c^\dagger \hat{U}_\theta^\dagger) \Omega \hat{a}^\dagger \hat{a} (\hat{U}_\theta \hat{U}_c) |\alpha\rangle$, where $\Omega$ denotes the bosonic mode frequency. In addition, both schemes are constrained to operate with the same total sensing time $T = t_c + t_\theta$. This condition leads to $|\alpha_0|^2 = \langle \alpha | (\hat{U}_c^\dagger \hat{U}_\theta^\dagger) \hat{a}^\dagger \hat{a} (\hat{U}_\theta \hat{U}_c) |\alpha\rangle $.
For nonvacuum probes $\alpha \neq 0$, the ratio $\mathcal{R}(\theta)$ reads as follows
\begin{equation}
    \mathcal{R}(\theta) = \frac{\mathcal{F}(\theta)}{4 (t_c + t_\theta)^2 \langle \alpha | (\hat{U}_c^\dagger \hat{U}_\theta^\dagger) \hat{a}^\dagger \hat{a} (\hat{U}_\theta \hat{U}_c) |\alpha\rangle},
   \label{eq4}
\end{equation}
If $\mathcal{R}(\theta) > 1$, critical-assisted noncommutative preparation yields a genuine sensitivity enhancement under equal energy and time resources. 
Because the energy constraint implies $|\alpha_0| > |\alpha|$, any $\mathcal{R}(\theta) > 1$ further indicates that the CANP achieves higher precision per initial excitation, establishing its superiority in terms of energy efficiency.

\textit{Quantifying noncommutativity by Wigner-Yanase skew information---}To quantitatively characterize the noncommutativity underlying the CANP protocol, we employ the Wigner–Yanase skew information, which provides a quantitative measure of the noncommutativity between a quantum state and an observable~\cite{Luo2003}. It is defined as
\begin{equation}
S = -\frac{1}{2} \text{Tr}([\sqrt{B},K]^2),
\label{eq5}
\end{equation}
where $B$ is a positive operator and $K$ is a Hermitian operator. The skew information $S$ captures the degree to which $K$ fails to commute with $B$, thereby serving as a natural probe of operator noncommutativity relevant to our scheme.

In our setting, we take $B = e^{-it_c\hat{H}_c}|\alpha\rangle \langle \alpha|e^{it_c\hat{H}_c} = \rho_c$, the probe state prepared through the criticality-assisted preparation governed by $\hat{H}_c$, while $K$ is identified with the encoding Hamiltonian $\hat{H}_\theta$. The resulting skew information thus quantifies the noncommutative degree between the encoding dynamics and the criticality-prepared state $\rho_c$.

Since $\rho_c$ is a pure state, the Wigner–Yanase skew information simplifies to the variance of the encoding Hamiltonian $\hat{H}_\theta$ in the prepared probe,
$
S = \text{Var}[\hat{H}_\theta]_{\rho_c} = \text{Var}[\hat{h}/t_\theta]_{\rho},
$
and thus connects directly to the QFI as
\begin{equation}
    S = \mathcal{F}(\theta)/(4 t_\theta^2) \simeq \frac{[\cos (\sqrt{\Delta} t_c)-1]^2}{\Delta^2} \text{Var}[\hat D]_{\rho}.
    \label{eq7}
\end{equation}
This correspondence shows that the skew information exhibits the same critical behavior as the QFI: it diverges as $\Delta \to 0$ for finite $t_c$ and shares the same oscillation structure set by the critical dynamics. Therefore, the Wigner–Yanase skew information provides a direct diagnostic of the intrinsic noncommutativity between the prepared probe state and the encoding operator,  identifying it as the fundamental resource responsible for the enhanced metrological sensitivity enabled by the CANP protocol.

\textit{An illustrative example: frequency estimation in the quantum Rabi model---}As an illustrative example of the CANP scheme, we take the quantum Rabi model as the critical preparation Hamiltonian, which is a paradigmatic system in quantum optics where a spin interacts with a single bosonic mode. The Hamiltonian, $\hat{H}_{\text{Rabi}} = \omega \hat{a}^\dagger \hat{a} + \frac{\omega_0}{2}\sigma_z^{(q)} - \lambda (\hat{a}+\hat{a}^\dagger)\sigma_x^{(q)}$, exhibits a normal-to-superradiant quantum phase transition, making it an ideal platform for criticality-assisted state preparation. Here, $\sigma_{x,z}^{(q)}$ denote Pauli operators of the qubit with transition frequency $\omega_0$, while $\hat{a}$ ($\hat{a}^\dagger$) are the annihilation (creation) operators of the field mode with frequency $\omega$, and $\lambda$ is the coupling parameter~\cite{Martin2015}.

In the normal phase, a Schrieffer–Wolff transformation yields the effective low-energy Hamiltonian,
\begin{equation}
\hat{H}_{\text{eff}} = \omega \hat{a}^\dagger \hat{a} - \frac{\omega g^2}{4} (\hat{a}^\dagger + \hat{a})^2 - \frac{\omega_0}{2},
\end{equation}
where $g = \frac{2\lambda}{\sqrt{\omega \omega_0}}$ is the renormalized coupling parameter. In the limit $\eta = \omega_0/\omega \to \infty$, the model undergoes a quantum phase transition at the critical point $g_c = 1$.

\begin{figure}
    \centering
    \includegraphics[width=1\linewidth]{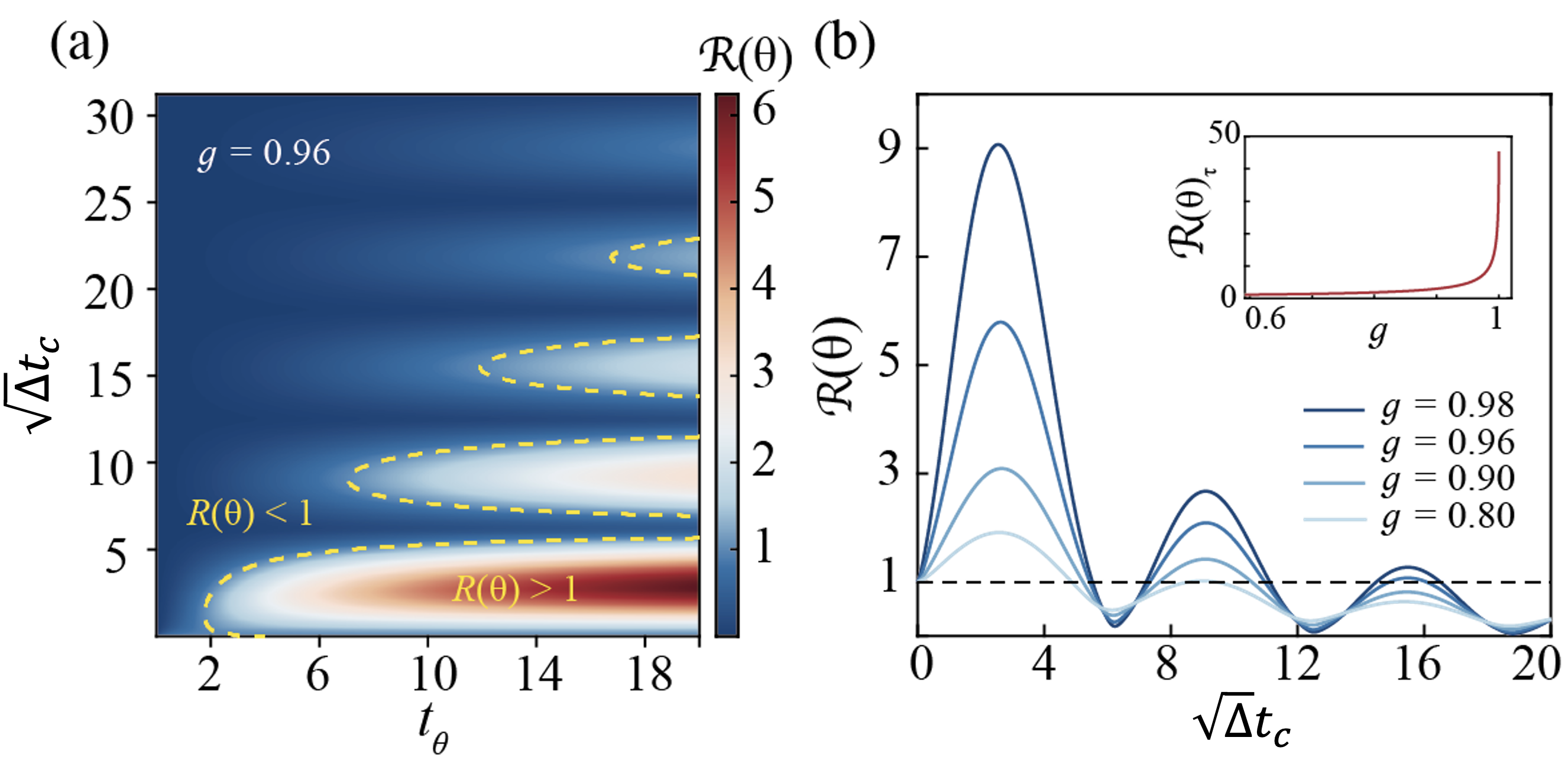}
    \caption{(a) QFI enhancement ratio $\mathcal{R}(\theta)$ as functions of the critical-evolution time $\sqrt{\Delta} t_c$ and encoding duration $t_\theta$. The probe amplitude is fixed at $\alpha = 0.3 + 1i$, and the critical parameter is set to $g=0.96$. The yellow dashed lines mark the reference value $\mathcal{R}(\theta) = 1$. (b) QFI enhancement ratio $\mathcal{R}(\theta)$ as functions of critical-evolution time $\sqrt{\Delta} t_c$ for different critical parameters $g = 0.80$, 0.90, 0.96 and 0.98. The black dashed line indicates $\mathcal{R}(\theta) = 1$. Other parameters are $t_\theta = 12$ and $\alpha = 0.3 + 1i$. Inset: the QFI enhancement ratio $\mathcal{R}(\theta)_\tau$ where $\tau = \pi/\sqrt{\Delta}$ for different $g$ values.}
    \label{fig2}
\end{figure}

We analyze a CANP protocol for frequency estimation implemented through
$
\hat{U} = \hat{U}_{\theta}\hat{U}_{c} =
e^{-it_\theta\theta \hat{H}_\theta}
e^{-it_c \hat{H}_{\text{eff}}},
$
where the parameter $\theta$ is encoded via $\hat{H}_\theta = \hat{a}^\dagger \hat{a}$.
Introducing $\Delta = 4 \omega^2 (1 - g^2)$, the noncommutative structure between $\hat{H}_{\text{eff}}$ and $\hat{H}_\theta$ is captured by the operators $\hat{C} = \frac{i \omega g^2}{2} ((\hat{a}^\dagger)^2 - \hat{a}^2)$ and $\hat{D} = g^2 \omega^2 [(1-g^2/2)((\hat{a}^\dagger)^2 + \hat{a}^2) - g^2 (\hat{a}^\dagger \hat{a} +1/2 )]$. Notice that while in principle the criticality parameter depends on frequency, we are not using this information. The critical evolution $U_c$ is used only to prepare the probe, and frequency information is encoded only by the unitary $U_\theta$.
When $g$ approaches to $g_c$ ($\Delta \to 0$), the dominant contribution of the QFI scales as
$
    \mathcal{F}(\theta) \simeq 4t_\theta^2 \Delta^{-2} [\cos (\sqrt{\Delta} t_c)-1]^2 \text{Var}[\hat D]_{\rho}.
$
For a finite preparation time $t_c \sim \mathcal{O}(1)$, this expression yields $\mathcal{F}(\theta) \sim t_\theta^2 t_c^4$, indicating an accelerated growth of sensitivity in the short-$t_c$ regime. By contrast, when the preparation time is sufficiently long and satisfies $t_c \sim \pi/\sqrt{\Delta}$, the QFI exhibits a divergent scaling $\mathcal{F}(\theta) \sim 16 \Delta^{-2} t_\theta^2 $ as the system approaches the critical point.

Moreover, to fairly benchmark the performance of the CANP protocol scheme against the conventional direct-encoding scheme $\hat{U}_\theta = e^{-i t_\theta \theta \hat{H}_\theta}$, we evaluate the enhancement ratio $\mathcal{R}(\theta)$ subject to the energy and total sensing time constraints (see details in Supplemental Material~\cite{sm}).
Figure~\ref{fig2}(a) displays $\mathcal{R}(\theta)$ as a function of $\sqrt{\Delta}t_c$, the time required 
for critical preparation, and $t_\theta$, the encoding time, for a fixed renormalized coupling parameter $g = 0.96$. The genuine enhancement emerges within several windows of the preparation time $t_c$ and encoding 
time $t_\theta$. Figure~\ref{fig2}(b) shows the dependence of $\mathcal{R}(\theta)$ on  $\sqrt{\Delta}t_c$ 
for fixed $t_\theta$. It can be seen that as $\sqrt{\Delta}t_c$ increases, the peaks of the oscillations in 
$\mathcal{R}(\theta)$ gradually diminish and eventually vanish. 
This behavior reflects a fundamental trade-off between the preparation and encoding stages: while the CANP QFI oscillates with the preparation time, the direct-encoding QFI grows quadratically with the total sensing duration. For sufficiently large $t_c$, this quadratic growth eventually overtakes the oscillation contribution in the CANP scheme, thereby suppressing the enhancement. 

Furthermore, the inset displays $\mathcal{R}(\theta)_\tau$ evaluated at $t_c = \tau = \pi/\sqrt{\Delta}$, as a function of $g$. The enhancement grows markedly as the system approaches criticality, with $\mathcal{R}(\theta)$ diverging in the limit $g \to g_c$. 
Such divergence reflects a generic feature of criticality-based quantum sensing, which is attainable for sufficiently long preparation times (i.e., $\tau \gg 1$). Nevertheless, numerical analysis shows that $\mathcal{R}(\theta) > 1$ already for $g > 0.5058$, demonstrating that the CANP protocol provides a genuine enhancement well before reaching the critical regime, and that the chosen preparation time $\tau = \pi/\sqrt{\Delta}$ remains finite. Therefore, the enhancement ratio $\mathcal{R}(\theta)$ provides a quantitative criterion for balancing the precision gain against the time cost of the overall evolution.

\begin{figure}
    \centering
    \includegraphics[width=1\linewidth]{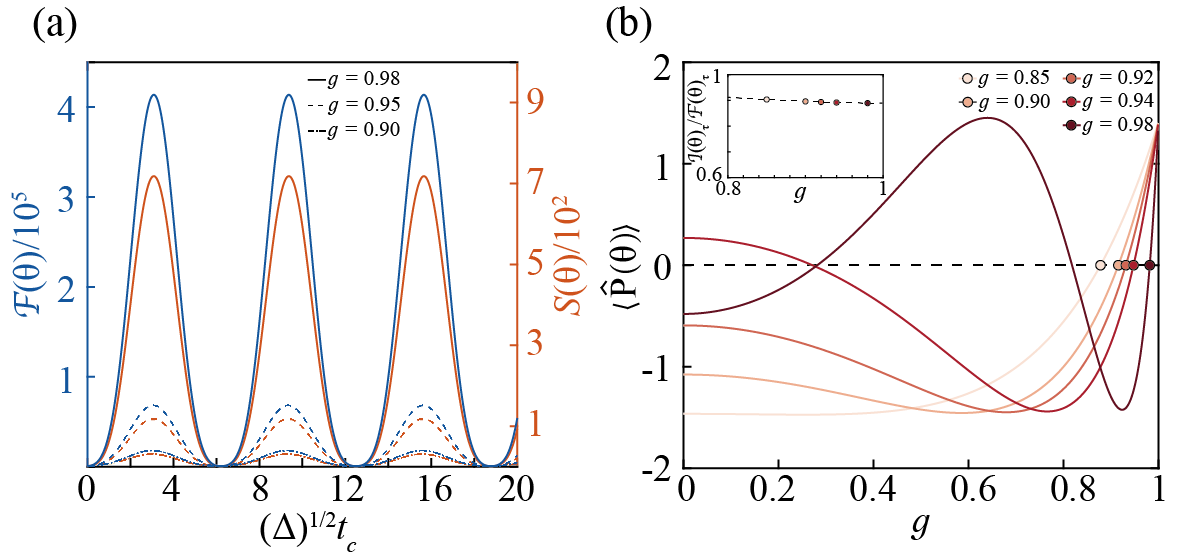}
    \caption{(a) Wigner–Yanase skew information $S(\theta)$ and the QFI $\mathcal{F}(\theta)$ as functions of the critical-evolution duration $\sqrt{\Delta} t_c$ for $g = 0.90$, 0.95 and 0.98, with an encoding evolution time fixed at $t_\theta$ = 12.  (b) The expectation value of the momentum quadrature $\langle \hat{P}(\theta) \rangle_{\tau}$ after a critical-evolution duration $\tau = \pi/\sqrt{\Delta}$ as a function of $g$. The working point $\langle \hat{P}(\theta) \rangle_{\tau} = 0$ is indicated by filled circles. Inset: the ratio of the classical Fisher information $\mathcal{I}(\theta)_\tau$ for quadrature measurement $\hat{P}$ to the QFI $\mathcal{F}(\theta)_\tau$ after a critical-evolution duration $\tau = \pi/\sqrt{\Delta}$. Other parameters are $t_\theta = 12$ and $\theta = 0$.}
    \label{fig3}
\end{figure}

To further characterize the noncommutativity between the effective Rabi Hamiltonian $\hat{H}_{\text{eff}}$ and the encoding Hamiltonian $\hat{H}_\theta$, we evaluate the Wigner–Yanase skew information $S(\theta)$ as shown in Fig.~\ref{fig3}(a). The skew information exhibits simultaneous oscillations and shares the critical point with the QFI, reflecting that the metrological enhancement is rooted in the intrinsic noncommutative structure of the dynamics.

Finally, we recall that the attainable precision is bounded by the quantum Cramér-Rao bound, determined by the inverse of the QFI and achievable only with optimal measurements. We hereby provide a feasible measurement strategy that achieves a precision scaling of the same order as this ultimate bound.

\textit{Measurement strategy for QRM-preparation sensing.---}
A natural measurement strategy for the bosonic mode is standard homodyne detection of a field quadrature. Without loss of generality, we consider measurements of the momentum quadrature $\hat{P} = i(\hat{a}^\dagger -\hat{a})/\sqrt{2}$. As shown in Fig.~\ref{fig3}(b), the expectation value $\langle \hat{P}(\theta)\rangle_\tau$ develops an increasingly steep slope near the working point $\langle \hat{P}(\theta)\rangle_\tau = 0$ with $t_c = \tau = \pi/\sqrt{\Delta}$ as the system approaches criticality, indicating enhanced sensitivity of the measurement signal.
Since both the preparation and the encoding stages preserve Gaussianity, the corresponding classical Fisher information (FI) can be evaluated analytically as $\mathcal{I}(\theta) = \frac{(\partial_\theta \langle \hat{P} \rangle)^2}{\Delta^2 \hat{P}} + \frac{1}{2} \frac{(\partial_\theta \Delta^2 \hat{P})^2}{(\Delta^2 \hat{P})^2}$, where $\Delta^2 \hat{P}$ denotes the variance of $\hat{P}$. The ratio $\mathcal{I}(\theta)_\tau / \mathcal{F}(\theta)_\tau$ at $\tau = \pi/\sqrt{\Delta}$, shown in the inset of Fig.~\ref{fig3}(b) for various values of $g$, reveals that the classical FI reaches approximately $90\%$ of the QFI across the parameter range. It can be seen that $\mathcal{I}(\theta)_\tau$ approaches the same order of magnitude as QFI even near the critical point. In addition, the signal-to-noise ratio associated with measurement of $\hat{P}$ remains robust under the beam-splitter loss model (see details in Supplemental Material~\cite{sm}). These results demonstrate that simple homodyne detection can attain a precision scaling comparable to the quantum Cram\'er-Rao limit.

\textit{Other Example: momentum-displacement estimation---}Within the CANP protocol, the choice of unknown parameters will not restrict the class of viable estimation protocols. This flexibility allows us to extend the framework to displacement estimation. Specifically, we take the noncritical generator $\hat{H}_p = (\hat{a}^\dagger + \hat{a})/\sqrt{2}$ to encode the momentum displacement $p$. Notably, $p$ does not appear in the critical Hamiltonian $\hat{H}_{\text{eff}}$ and is therefore inaccessible within the conventional critical metrology schemes based on the QRM.
To estimate the displacement parameter $p$, we consider the CANP protocol $\hat{U} = \exp(-ipt_p \hat{H}_p)\exp(-it_c \hat{H}_{\text{eff}})$. The critical point of the effective Rabi dynamics, $g_c = 1$, corresponds to $\Delta_p = \omega^2 (1-g^2) = 0$. Since its noncommutativity between $\hat{H}_p$ and $\hat{H}_{\text{eff}}$ satisfies the criterion in Eq.~({\ref{eq1}}) and following the same procedure as in the general framework, we obtain the QFI $\mathcal{F}(p) \simeq 4t_p^2 \omega^2 \Delta_p^{-1} \sin^2(\sqrt{\Delta_p}t_c) \text{Var}[\hat{P}]_\rho$, which exhibits critical enhancement as $g \to g_c$ for any initial state $\rho$ with nonzero momentum variance.
Consistently, the enhancement ratio $\mathcal{R}(p)$ displays a substantive enhancement in the critical limit with fix energy cost and total sensing time (See details in Supplemental Material~\cite{sm}), further demonstrating that CANP generically boosts displacement sensing.

\textit{Conclusion---}In this Letter, we have presented CANP, a protocol that leverages critical quantum dynamics as a state-preparation resource for quantum metrology, without requiring additional energy or total sensing time. In particular, we have shown that a genuine enhancement of the QFI is achievable by exploiting the noncommutativity, as quantified the Wigner–Yanase skew information, between the critical preparation and the parameter-encoding generator.  We have illustrated the principle by taking the QRM as an explicit example of critical preparation for frequency estimation, and shown that the analysis and the results can be extended to more examples, such as the Lipkin-Meshkov-Glick model (see details in End Matter).

Our results show that relocating the critical Hamiltonian from the encoding stage to the preparation stage naturally enables the exploitation of criticality for estimating parameters generated by noncritical operators, and eliminating the narrow effective estimation range imposed by the requirement of critical conditions.
In addition to the quantum-phase-transition-induced criticality discussed in the main text, this framework extends straightforwardly to scenarios with tunable forms of criticality, including the non-Hermitian exceptional point dynamics~\cite{Du2022,Zhang2024,He2025b}, as detailed in the Supplementary Material~\cite{sm}.
Overall, the CANP protocol demonstrates that the combination of criticality and noncommutativity offers a powerful and resource-efficient route to quantum enhancement. This mechanism opens a pathway for better incorporating critical phenomena into metrological tasks, extends the advantages of the conventional critical-encoding setting, and thereby overcomes limitations in current critical metrology.

\textit{Acknowledgments---}This work was supported by Quantum Science and Technology-National Science and Technology Major Project (Grant No. 2024ZD0302401 and No. 2021ZD0301500), National Natural Science Foundation of China (No. 12125402 and No. 12534016), and Beijing Natural Science Foundation (Grant No. Z240007). 

\textit{End Matter---}The principles underlying the CANP protocol extend broadly to Hamiltonians whose algebraic structure satisfies the Eq.~(\ref{eq1}). As a second example, we consider the Lipkin-Meshkov-Glick (LMG) model. Applying the Holstein-Primakoff transformation and in the low-excitation regime $\langle \hat{a}^\dagger \hat{a}\rangle \ll N$, the effective LMG Hamiltonian can be approximated as $\hat{H}_{\text{LMG}} = 2\lambda \hat{a}^\dagger \hat{a} + [\gamma(\hat{a}^\dagger - \hat{a})^2 - (\hat{a} + \hat{a}^\dagger)^2]/2$. One verifies that its algebraic structure with the encoding Hamiltonian $\hat{H}_\theta = \hat{a}^\dagger \hat{a}$ satisfies Eq.~(\ref{eq1}), yielding the critical parameter $\Delta_\lambda = 16(\gamma - \lambda)(1-\lambda)$. For $\gamma \neq 1$, the model exhibits a quantum phase transition at the critical point $\lambda_c =1$.
Following the same analytical procedure used for the QRM, we find that the QFI scales as $\mathcal{F}(\theta) \simeq 4 t_\theta^2 \Delta_\lambda^{-2} [\cos(\sqrt{\Delta_\lambda}t_c)-1]^2 \text{Var}[\hat{D}]_\rho$, where $\hat{D} = 2(\gamma - 1)[(1+\gamma -2\lambda)((\hat{a}^\dagger)^2 + \hat{a}^2) + (1-\gamma)(2\hat{a}^\dagger \hat{a}+1)]$. The QFI enhancement ratio already satisfies $\mathcal{R}(\theta)_\tau > 1$ for $\lambda > 0.3559$ with $t_\theta = 1.3$ and $\gamma = 2$, far from the critical regime. This demonstrates that the CANP protocol displays a genuine enhancement even for short preparation durations $\tau = \pi/\sqrt{\Delta_{\lambda}}$. As the system approaches the critical point $\lambda \to \lambda_c$, the enhancement becomes increasingly pronounced, and $\mathcal{R}(\theta)$ diverges (See details in Supplemental Material~\cite{sm}).

\bibliography{ref_critical}

@misc{sm,
	note = {See Supplemental Material for detailed demonstrations of the criticality-assisted non-commutative preparation (CANP) protocol applied to frequency estimation in the quantum Rabi model together with the corresponding measurement scheme, frequency estimation in the Lipkin–Meshkov–Glick model, momentum-displacement estimation in the quantum Rabi model, as well as implementations of the CANP protocol based on exceptional-point-induced criticality in non-Hermitian dynamics}}

@article{Zhang2024,
  title={Scaling of quantum Fisher information for quantum exceptional point sensors},
  author={ Liu, Chun-Hui and Li, Fu and Du, Shengwang and Wen, Jianming and Yang, Lan and Zhang, Chuanwei},
  journal={arXiv:2404.03803},
  year={2024},
  url       = {https://arxiv.org/abs/2404.03803},
}

@article{Du2022,
  title = {Quantum Squeezing and Sensing with Pseudo-Anti-Parity-Time Symmetry},
  author = {Luo, Xi-Wang and Zhang, Chuanwei and Du, Shengwang},
  journal = {Phys. Rev. Lett.},
  volume = {128},
  issue = {17},
  pages = {173602},
  numpages = {7},
  year = {2022},
  month = {Apr},
  publisher = {American Physical Society},
  doi = {10.1103/PhysRevLett.128.173602},
  url = {https://link.aps.org/doi/10.1103/PhysRevLett.128.173602}
}

@article{He2025b,
	author = {Yu, Chenghe and Tian, Mingsheng and Kong, Ningxin and Fadel, Matteo and Huang, Xinyao and He, Qiongyi},
	doi = {10.1038/s41534-025-01158-y},
	isbn = {2056-6387},
	journal = {npj Quantum Inf.},
	number = {1},
	pages = {14},
	title = {Exceptional-point-induced nonequilibrium entanglement dynamics in bosonic networks},
	url = {https://doi.org/10.1038/s41534-025-01158-y},
	volume = {12},
	year = {2025}}

@article{Maccone2004,
        author = {Vittorio Giovannetti  and Seth Lloyd  and Lorenzo Maccone },
        title = {Quantum-Enhanced Measurements: Beating the Standard Quantum Limit},
        journal = {Science},
        volume = {306},
        number = {5700},
        pages = {1330-1336},
        year = {2004},
        doi = {10.1126/science.1104149},
        URL = {https://www.science.org/doi/abs/10.1126/science.1104149}
}

@article{Maccone2006,
  title = {Quantum Metrology},
  author = {Giovannetti, Vittorio and Lloyd, Seth and Maccone, Lorenzo},
  journal = {Phys. Rev. Lett.},
  volume = {96},
  issue = {1},
  pages = {010401},
  numpages = {4},
  year = {2006},
  month = {Jan},
  publisher = {American Physical Society},
  doi = {10.1103/PhysRevLett.96.010401},
  url = {https://link.aps.org/doi/10.1103/PhysRevLett.96.010401}
}

@article{Romalis2007,
	title = {Optical magnetometry},
	volume = {3},
	issn = {1745-2473},
	url = {https://www.nature.com/articles/nphys566},
	doi = {10.1038/nphys566},
	number = {4},
	journal = {Nat. Phys.},
	author = {Budker, Dmitry and Romalis, Michael},
	month = {Apr},
	year = {2007},
	pages = {227--234}
}

@article{PARIS2009,
author = {Paris, Matteo G. A.},
title = {Quantum estimation for quantum technology},
journal = {Int. J. Quantum Inf.},
volume = {07},
pages = {125-137},
year = {2009},
doi = {10.1142/S0219749909004839},
URL = {https://doi.org/10.1142/S0219749909004839}
}

@article{Cappellaro2017,
	title = {Quantum sensing},
	volume = {89},
	copyright = {http://link.aps.org/licenses/aps-default-license},
	issn = {0034-6861},
	url = {http://link.aps.org/doi/10.1103/RevModPhys.89.035002},
	doi = {10.1103/RevModPhys.89.035002},
	number = {3},
	urldate = {2024-07-17},
	journal = {Rev. Mod. Phys.},
	author = {Degen, C. L. and Reinhard, F. and Cappellaro, P.},
	month = {Jul},
	year = {2017},
	pages = {035002}
}

@article{Treutlein2018,
  title = {Quantum metrology with nonclassical states of atomic ensembles},
  author = {Pezz\`e, Luca and Smerzi, Augusto and Oberthaler, Markus K. and Schmied, Roman and Treutlein, Philipp},
  journal = {Rev. Mod. Phys.},
  volume = {90},
  issue = {3},
  pages = {035005},
  numpages = {70},
  year = {2018},
  month = {Sep},
  publisher = {American Physical Society},
  doi = {10.1103/RevModPhys.90.035005},
  url = {https://link.aps.org/doi/10.1103/RevModPhys.90.035005}
}

@article{Maccone2011,
	title = {Advances in quantum metrology},
	volume = {5},
	copyright = {http://www.springer.com/tdm},
	issn = {1749-4885},
	url = {https://www.nature.com/articles/nphoton.2011.35},
	doi = {10.1038/nphoton.2011.35},
	number = {4},
	urldate = {2024-07-17},
	journal = {Nat. Photon.},
	author = {Giovannetti, Vittorio and Lloyd, Seth and Maccone, Lorenzo},
	month = {Apr},
	year = {2011},
	pages = {222--229}
}

@article{Mitchell2011,
  title={Interaction-based quantum metrology showing scaling beyond the Heisenberg limit},
  author={Napolitano, Mario and Koschorreck, Marco and Dubost, Brice and Behbood, Naeimeh and Sewell, RJ and Mitchell, Morgan W},
  journal={Nature},
  volume={471},
  number={7339},
  pages={486--489},
  year={2011},
  publisher={Nature Publishing Group UK London},
  url = {https://www.nature.com/articles/nature09778}
}

@article{Zoller2024,
	title = {Essay: {Quantum} {Sensing} with {Atomic}, {Molecular}, and {Optical} {Platforms} for {Fundamental} {Physics}},
	volume = {132},
	issn = {0031-9007},
	url = {https://link.aps.org/doi/10.1103/PhysRevLett.132.190001},
	doi = {10.1103/PhysRevLett.132.190001},
	number = {19},
	urldate = {2024-09-19},
	journal = {Phys. Rev. Lett.},
	author = {Ye, Jun and Zoller, Peter},
	month = {May},
	year = {2024},
	pages = {190001}
}

@article{Zelevinsky2024,
	title = {Quantum sensing and metrology for fundamental physics with molecules},
	volume = {20},
	issn = {1745-2473},
	url = {https://www.nature.com/articles/s41567-024-02499-9},
	doi = {10.1038/s41567-024-02499-9},
	number = {5},
	urldate = {2024-07-17},
	journal = {Nat. Phys.},
	author = {DeMille, David and Hutzler, Nicholas R. and Rey, Ana Maria and Zelevinsky, Tanya},
	month = {may},
	year = {2024},
	pages = {741--749}
}

@article{Nikola2006,
  title = {Ground state overlap and quantum phase transitions},
  author = {Zanardi, Paolo and Paunkovi\ifmmode \acute{c}\else \'{c}\fi{}, Nikola},
  journal = {Phys. Rev. E},
  volume = {74},
  issue = {3},
  pages = {031123},
  numpages = {6},
  year = {2006},
  month = {Sep},
  publisher = {American Physical Society},
  doi = {10.1103/PhysRevE.74.031123},
  url = {https://link.aps.org/doi/10.1103/PhysRevE.74.031123}
}

@article{Lorenzo2008,
  title = {Quantum criticality as a resource for quantum estimation},
  author = {Zanardi, Paolo and Paris, Matteo G. A. and Campos Venuti, Lorenzo},
  journal = {Phys. Rev. A},
  volume = {78},
  issue = {4},
  pages = {042105},
  numpages = {7},
  year = {2008},
  month = {Oct},
  publisher = {American Physical Society},
  doi = {10.1103/PhysRevA.78.042105},
  url = {https://link.aps.org/doi/10.1103/PhysRevA.78.042105}
}

@article{Kehrein2013,
  title = {Dynamical Quantum Phase Transitions in the Transverse-Field Ising Model},
  author = {Heyl, M. and Polkovnikov, A. and Kehrein, S.},
  journal = {Phys. Rev. Lett.},
  volume = {110},
  issue = {13},
  pages = {135704},
  numpages = {5},
  year = {2013},
  month = {Mar},
  publisher = {American Physical Society},
  doi = {10.1103/PhysRevLett.110.135704},
  url = {https://link.aps.org/doi/10.1103/PhysRevLett.110.135704}
}

@article{Paris2016,
  title = {Dicke coupling by feasible local measurements at the superradiant quantum phase transition},
  author = {Bina, M. and Amelio, I. and Paris, M. G. A.},
  journal = {Phys. Rev. E},
  volume = {93},
  issue = {5},
  pages = {052118},
  numpages = {10},
  year = {2016},
  month = {May},
  publisher = {American Physical Society},
  doi = {10.1103/PhysRevE.93.052118},
  url = {https://link.aps.org/doi/10.1103/PhysRevE.93.052118}
}

@article{Tommaso2018,
  title = {Quantum Critical Metrology},
  author = {Fr\'erot, Ir\'en\'ee and Roscilde, Tommaso},
  journal = {Phys. Rev. Lett.},
  volume = {121},
  issue = {2},
  pages = {020402},
  numpages = {6},
  year = {2018},
  month = {Jul},
  publisher = {American Physical Society},
  doi = {10.1103/PhysRevLett.121.020402},
  url = {https://link.aps.org/doi/10.1103/PhysRevLett.121.020402}
}

@article{Nigel2019,
  title = {Critical Response of a Quantum van der Pol Oscillator},
  author = {Dutta, Shovan and Cooper, Nigel R.},
  journal = {Phys. Rev. Lett.},
  volume = {123},
  issue = {25},
  pages = {250401},
  numpages = {6},
  year = {2019},
  month = {Dec},
  publisher = {American Physical Society},
  doi = {10.1103/PhysRevLett.123.250401},
  url = {https://link.aps.org/doi/10.1103/PhysRevLett.123.250401}
}

@article{Bayat2025,
title = {Review: Quantum metrology and sensing with many-body systems},
journal = {Phys. Rep.},
volume = {1134},
pages = {1-62},
year = {2025},
issn = {0370-1573},
doi = {https://doi.org/10.1016/j.physrep.2025.05.005},
url = {https://www.sciencedirect.com/science/article/pii/S0370157325001565},
author = {Victor Montenegro and Chiranjib Mukhopadhyay and Rozhin Yousefjani and Saubhik Sarkar and Utkarsh Mishra and Matteo G.A. Paris and Abolfazl Bayat},
}

@article{Abolfazl2021,
  title = {Global Sensing and Its Impact for Quantum Many-Body Probes with Criticality},
  author = {Montenegro, Victor and Mishra, Utkarsh and Bayat, Abolfazl},
  journal = {Phys. Rev. Lett.},
  volume = {126},
  issue = {20},
  pages = {200501},
  numpages = {6},
  year = {2021},
  month = {May},
  publisher = {American Physical Society},
  doi = {10.1103/PhysRevLett.126.200501},
  url = {https://link.aps.org/doi/10.1103/PhysRevLett.126.200501}
}

@article{Simone2020,
  title = {Critical Quantum Metrology with a Finite-Component Quantum Phase Transition},
  author = {Garbe, Louis and Bina, Matteo and Keller, Arne and Paris, Matteo G. A. and Felicetti, Simone},
  journal = {Phys. Rev. Lett.},
  volume = {124},
  issue = {12},
  pages = {120504},
  numpages = {5},
  year = {2020},
  month = {Mar},
  publisher = {American Physical Society},
  doi = {10.1103/PhysRevLett.124.120504},
  url = {https://link.aps.org/doi/10.1103/PhysRevLett.124.120504}
}

@article{Cai2021,
  title = {Dynamic Framework for Criticality-Enhanced Quantum Sensing},
  author = {Chu, Yaoming and Zhang, Shaoliang and Yu, Baiyi and Cai, Jianming},
  journal = {Phys. Rev. Lett.},
  volume = {126},
  issue = {1},
  pages = {010502},
  numpages = {7},
  year = {2021},
  month = {Jan},
  publisher = {American Physical Society},
  doi = {10.1103/PhysRevLett.126.010502},
  url = {https://link.aps.org/doi/10.1103/PhysRevLett.126.010502}
}

@article{Karol2024,
  title = {Combining Critical and Quantum Metrology},
  author = {Hotter, Christoph and Ritsch, Helmut and Gietka, Karol},
  journal = {Phys. Rev. Lett.},
  volume = {132},
  issue = {6},
  pages = {060801},
  numpages = {8},
  year = {2024},
  month = {Feb},
  publisher = {American Physical Society},
  doi = {10.1103/PhysRevLett.132.060801},
  url = {https://link.aps.org/doi/10.1103/PhysRevLett.132.060801}
}

@article{Candia2024,
  title = {Optimality and Noise Resilience of Critical Quantum Sensing},
  author = {Alushi, U. and G\'orecki, W. and Felicetti, S. and Di Candia, R.},
  journal = {Phys. Rev. Lett.},
  volume = {133},
  issue = {4},
  pages = {040801},
  numpages = {7},
  year = {2024},
  month = {Jul},
  publisher = {American Physical Society},
  doi = {10.1103/PhysRevLett.133.040801},
  url = {https://link.aps.org/doi/10.1103/PhysRevLett.133.040801}
}

@article{Zeng2025,
  title = {Toward Heisenberg Limit without Critical Slowing Down via Quantum Reinforcement Learning},
  author = {Xu, Hang and Xiao, Tailong and Huang, Jingzheng and He, Ming and Fan, Jianping and Zeng, Guihua},
  journal = {Phys. Rev. Lett.},
  volume = {134},
  issue = {12},
  pages = {120803},
  numpages = {6},
  year = {2025},
  month = {Mar},
  publisher = {American Physical Society},
  doi = {10.1103/PhysRevLett.134.120803},
  url = {https://link.aps.org/doi/10.1103/PhysRevLett.134.120803}
}

@article{Jakub2018,
  title = {At the Limits of Criticality-Based Quantum Metrology: Apparent Super-Heisenberg Scaling Revisited},
  author = {Rams, Marek M. and Sierant, Piotr and Dutta, Omyoti and Horodecki, Pawe\l{} and Zakrzewski, Jakub},
  journal = {Phys. Rev. X},
  volume = {8},
  issue = {2},
  pages = {021022},
  numpages = {16},
  year = {2018},
  month = {Apr},
  publisher = {American Physical Society},
  doi = {10.1103/PhysRevX.8.021022},
  url = {https://link.aps.org/doi/10.1103/PhysRevX.8.021022}
}

@article{Giulio2020,
	title = {Quantum {Metrology} with {Indefinite} {Causal} {Order}},
	volume = {124},
	issn = {0031-9007},
	url = {https://link.aps.org/doi/10.1103/PhysRevLett.124.190503},
	doi = {10.1103/PhysRevLett.124.190503},
	number = {19},
	urldate = {2024-11-28},
	journal = {Phys. Rev. Lett.},
	author = {Zhao, Xiaobin and Yang, Yuxiang and Chiribella, Giulio},
	month = {May},
	year = {2020},
	pages = {190503}
}

@article{Guo2023,
	title = {Experimental super-{Heisenberg} quantum metrology with indefinite gate order},
	volume = {19},
	issn = {1745-2473},
	url = {https://www.nature.com/articles/s41567-023-02046-y},
	doi = {10.1038/s41567-023-02046-y},
	number = {8},
	urldate = {2024-11-28},
	journal = {Nat. Phys.},
	author = {Yin, Peng and Zhao, Xiaobin and Yang, Yuxiang and Guo, Yu and Zhang, Wen-Hao and Li, Gong-Chu and Han, Yong-Jian and Liu, Bi-Heng and Xu, Jin-Shi and Chiribella, Giulio and Chen, Geng and Li, Chuan-Feng and Guo, Guang-Can},
	month = {Aug},
	year = {2023},
	pages = {1122--1127}
}

@article{Zeng2024,
	title = {Nanoradian-scale precision in light rotation measurement via indefinite quantum dynamics},
	volume = {10},
	issn = {2375-2548},
	url = {https://www.science.org/doi/10.1126/sciadv.adm8524},
	doi = {10.1126/sciadv.adm8524},
	number = {28},
	urldate = {2024-11-28},
	journal = {Sci. Adv.},
	author = {Xia, Binke and Huang, Jingzheng and Li, Hongjing and Luo, Zhongyuan and Zeng, Guihua},
	month = {Jul},
	year = {2024},
	pages = {eadm8524},
}

@article{Wang2025,
  title = {Generalized Indefinite Causal Orders in an Integrated Quantum Switch},
  author = {Deng, Yaohao and Liu, Shuheng and Chen, Xiaojiong and Fu, Zhaorong and Bao, Jueming and Zheng, Yun and Gong, Qihuang and He, Qiongyi and Wang, Jianwei},
  journal = {Phys. Rev. Lett.},
  volume = {135},
  issue = {16},
  pages = {160202},
  numpages = {7},
  year = {2025},
  month = {Oct},
  publisher = {American Physical Society},
  doi = {10.1103/39vh-84n1},
  url = {https://link.aps.org/doi/10.1103/39vh-84n1}
}

@article{Paris2025,
  title = {Enhanced Quantum Frequency Estimation by Nonlinear Scrambling},
  author = {Montenegro, Victor and Dornetti, Sara and Ferraro, Alessandro and Paris, Matteo G. A.},
  journal = {Phys. Rev. Lett.},
  volume = {135},
  issue = {3},
  pages = {030802},
  numpages = {8},
  year = {2025},
  month = {Jul},
  publisher = {American Physical Society},
  doi = {10.1103/39bt-37yl},
  url = {https://link.aps.org/doi/10.1103/39bt-37yl}
}

@article{He2026,
  title = {Noncommutativity as a Universal Characterization for Enhanced Quantum Metrology},
  author = {Kong, Ningxin and Wang, Haojie and Tian, Mingsheng and Xu, Yilun and Chen, Geng and Xiang, Yu and He, Qiongyi},
  journal = {Phys. Rev. Lett.},
  volume = {136},
  issue = {1},
  pages = {010201},
  numpages = {6},
  year = {2026},
  month = {Jan},
  publisher = {American Physical Society},
  doi = {10.1103/3jlc-lb5c},
  url = {https://link.aps.org/doi/10.1103/3jlc-lb5c}
}

@article{Zeng2025b,
  title={Nonlinear Heisenberg limit via uncertainty principle in quantum metrology},
  author={Xia, Binke and Huang, Jingzheng and Yang, Yuxiang and Zeng, Guihua},
  journal={arXiv:2510.09216},
  year={2025},
  url       = {https://arxiv.org/abs/2510.09216v1},
}

@article{Caves1994,
  title = {Statistical distance and the geometry of quantum states},
  author = {Braunstein, Samuel L. and Caves, Carlton M.},
  journal = {Phys. Rev. Lett.},
  volume = {72},
  issue = {22},
  pages = {3439--3443},
  numpages = {0},
  year = {1994},
  month = {May},
  publisher = {American Physical Society},
  doi = {10.1103/PhysRevLett.72.3439},
  url = {https://link.aps.org/doi/10.1103/PhysRevLett.72.3439}
}

@article{Kok2010,
	title = {General {Optimality} of the {Heisenberg} {Limit} for {Quantum} {Metrology}},
	volume = {105},
	copyright = {http://link.aps.org/licenses/aps-default-license},
	issn = {0031-9007},
	url = {https://link.aps.org/doi/10.1103/PhysRevLett.105.180402},
	doi = {10.1103/PhysRevLett.105.180402},
	number = {18},
	urldate = {2025-02-09},
	journal = {Phys. Rev. Lett.},
	author = {Zwierz, Marcin and Pérez-Delgado, Carlos A. and Kok, Pieter},
	month = {Oct},
	year = {2010},
	pages = {180402},
}

@article{Luo2003,
  title = {Wigner-Yanase Skew Information and Uncertainty Relations},
  author = {Luo, Shunlong},
  journal = {Phys. Rev. Lett.},
  volume = {91},
  issue = {18},
  pages = {180403},
  numpages = {4},
  year = {2003},
  month = {Oct},
  publisher = {American Physical Society},
  doi = {10.1103/PhysRevLett.91.180403},
  url = {https://link.aps.org/doi/10.1103/PhysRevLett.91.180403}
}

@article{Martin2015,
  title = {Quantum Phase Transition and Universal Dynamics in the Rabi Model},
  author = {Hwang, Myung-Joong and Puebla, Ricardo and Plenio, Martin B.},
  journal = {Phys. Rev. Lett.},
  volume = {115},
  issue = {18},
  pages = {180404},
  numpages = {5},
  year = {2015},
  month = {Oct},
  publisher = {American Physical Society},
  doi = {10.1103/PhysRevLett.115.180404},
  url = {https://link.aps.org/doi/10.1103/PhysRevLett.115.180404}
}

\end{document}